\documentclass{acm_proc_article-sp}
\usepackage{graphicx}
\usepackage{textcomp}
\usepackage{fancyvrb}
\usepackage{listings}
\usepackage[unicode=true]{hyperref}
\usepackage{enumitem}

\begin{document}

\conferenceinfo{}{Bloomberg Data for Good Exchange 2016, NY, USA}

\title{"Birds in the Clouds": Adventures in Data Engineering}

\numberofauthors{2}
\author{
\alignauthor
N. Cherel, J. Reesman, A. Sahuguet\\
       \affaddr{The Foundry @ Cornell Tech}\\
       \affaddr{Cornell Tech, New York USA}\\
       \email{\{nac243,jhr265,arnaud.sahuguet\}@cornell.edu}
\and
T. Auer,  D. Fink\\
       \affaddr{Cornell Laboratory of Ornithology}\\
       \affaddr{Ithaca, NY}\\
       \email{\{daniel.fink,mta45\}@cornell.edu}
}

\maketitle \begin{abstract}
Leveraging their eBird crowdsourcing project, the Cornell Lab of Ornithology generates sophisticated Spatio-Temporal Exploratory Model (STEM) maps of bird migrations. Such maps are highly relevant for both scientific and educational purposes, but creating them requires advanced modeling techniques that rely on long and potentially expensive computations.
    
In this paper, we share our experience porting the eBird data pipeline from a physical cluster to the cloud, providing a seamless deployment at a lower cost. Using open source tools and cloud "marketplaces", we managed to divide the operating costs by a factor of 6, saving  hundreds of thousands of dollars.
\end{abstract}

\section{Introduction}\label{sec::intro}
Birds provide unrivalled insight into overall ecosystem health. They occur in almost all environments around the globe and their migratory patterns integrate disparate ecosystems across regional, continental, and hemispheric scales. Migratory birds are vulnerable to a number of environmental threats including habitat conversion and fragmentation, intensification of incompatible agriculture practices, unsuitable water management, water scarcity and drought, and climate change. This makes these populations valuable indicators of global change, but also points to the challenges that these populations face for their survival.

Assessing the health of bird populations -- and through them, the health of ecosystems -- requires a full lifecycle analysis, including breeding, migration, and wintering patterns. Since the lifecycle of many species of birds spans enormous spatial scales, includes many different habitats, and varies widely between years, obtaining the necessary information can be very challenging.

However, birds are easy to observe and identify, and, importantly, they have been objects of inspiration for people over the millennia. Today, rich observational data on bird populations is collected by bird-watching enthusiasts, birders, the world over. By providing tools to birders, the eBird project, run by the \href{http://www.birds.cornell.edu/}{Cornell Lab of Ornithology} (CLO), transforms a recreational activity into citizen science, where the public can play an active part in scientific research. Through the eBird project, millions of bird observations are being recorded, curated, published, and used for scientific research. Processing all these observations is the right thing to do, but it can be very expensive.

In this paper, we report our experience running the eBird STEM analytical pipeline. The pipeline was developed to transform observations from birders into migration maps for thousands of bird species. Initial runs of the pipeline on our local cluster cost hundreds of thousands of dollars. Given shrinking research budgets, this translated into less frequent runs or arbitrary choices about which species to prioritize. In this paper, we describe our "migration" (pun intended) to the clouds, and what we learned doing it.

The rest of the paper is organized as follows. We first present an overview of the eBird project and how crowdsourced data can be used to create migration models. We then give an overview of the eBird data pipeline and the intuition behind the computations it performs. In Section~\ref{sec::migration}, we describe our migration to the cloud. We share some lessons learned in Section~\ref{sec::lessons}.

Results presented in this paper are based our personal experience using the various cloud technologies at our disposal, to the best of our knowledge and abilities. Your mileage may vary.

\section{The eBird Project}\label{sec::ebird}
eBird, launched in 2002 by the Cornell Lab of Ornithology (CLO), is a citizen science project that uses crowdsourcing techniques to collect bird monitoring data around the globe throughout the year. Using a web interface or mobile app, bird watching enthusiasts or birders follow a simple protocol in which they collect observations of the bird species they see -- checklists -- along with valuable ancillary information about the time, location, search effort, and media recordings, if any. The eBird database contains observation for more than 10,313  bird species. This information is aggregated and curated for both scientific and educational uses.

\begin{figure}[h]
    \centering
    \includegraphics[width=8cm]{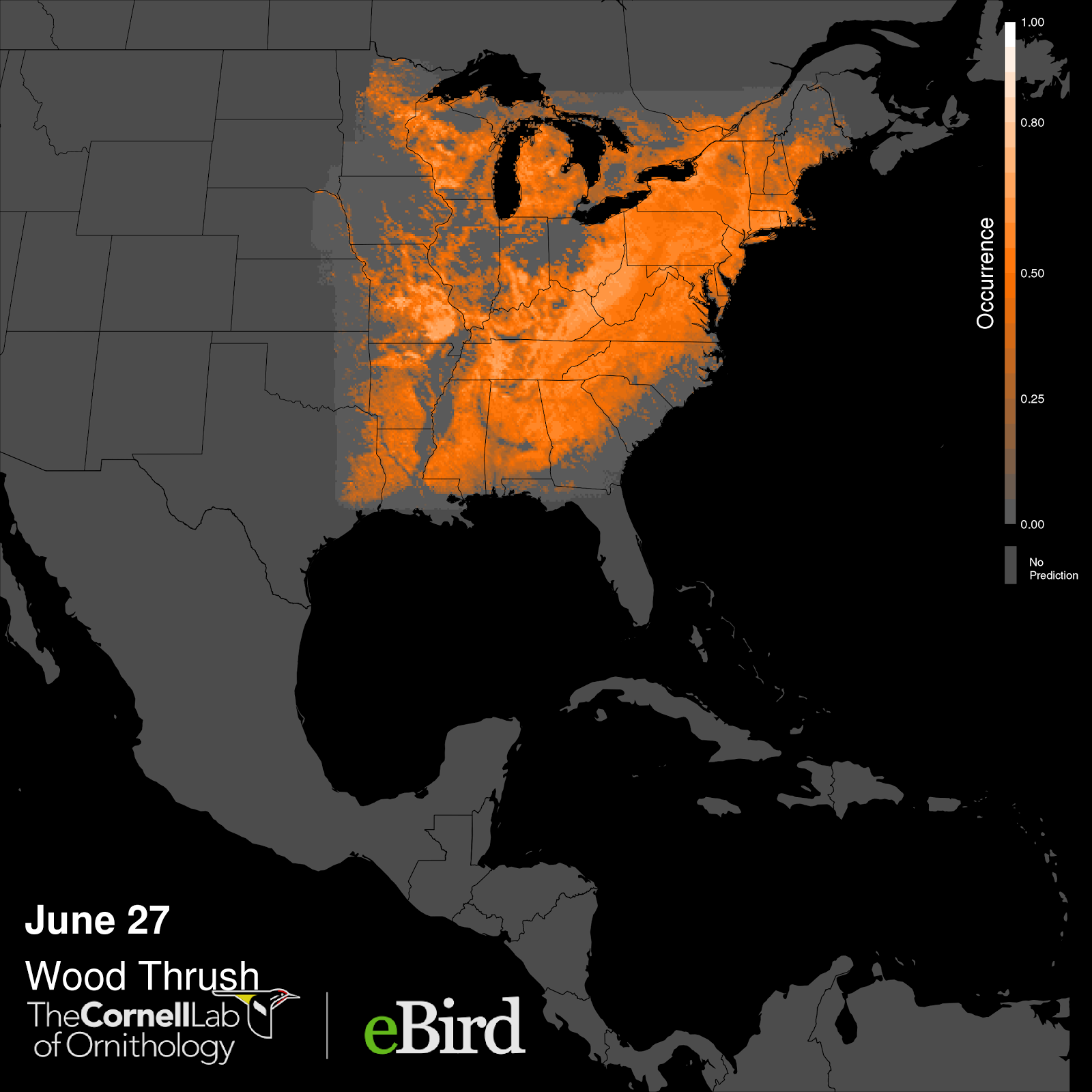}
    \caption{Occurrence map for Wood Thrush on June 27.}
    \label{fig::stem}
\end{figure}

One key research application of the eBird data is the creation of Spatio-Temporal Exploratory Model (STEM) maps that describe the migration of birds. An example of such a map is presented in Figure~\ref{fig::stem} and clearly showcases the power of eBird data and STEM for year-round, hemisphere-scale monitoring of migration for all species.

The logic behind the STEM modeling and the map creation can be summarized as follows. The location of each checklist is associated with remotely-sensed information on local land cover, water cover, and topography. This generates a suite of approximately 80 variables describing the environment where eBird searches take place. By relating these environmental variables to observed occurrences, STEM makes predictions at unsampled locations and times. 

Models are trained one species at a time. Following model training, the expected occurrence for that species is predicted on each of 52 days, one per week at some 1M locations sampled throughout the terrestrial Western Hemisphere\footnote{Western Hemisphere north to 72 degrees latitude.}. This massive volume of information is then summarized on maps, which in many cases reveal novel information about the annual cycles of these bird populations. These maps showcase the power of eBird -- year-round, hemisphere-scale monitoring of all species.

\section{The STEM analytical Pipeline}\label{sec::pipeline}
To generate STEM map data for a given species, the Cornell Lab of Ornithology needs to aggregate existing data and predict missing data for that species across a large geographic area for a year-long window. In this section, we provide a high level description of the computations required to build STEM maps. For more technical information, we point the reader to \cite{Fink2010-fb} and \cite{Johnston2015-pf}.

The eBird data currently used in the STEM workflow contains millions of search records for each species in space and time, collected year round and through all over the Western Hemisphere. A species distribution, however, is best explained at a local scale of space and time. This concept of location and time is encapsulated as a stixel, which represents a two-dimensional region for a given time window.

The data pipeline processes observations gathered from eBird crowdsourced data to make predictions about the presence (0 or 1) and abundance (counts) of a given species at a given time and a given location. An observation is defined by its timestamp; its location (latitude and longitude); a description of the environment where it was made (e.g. forest, lake, etc.); a description of the way the observation was conducted (e.g. search duration, time of day search conducted, etc.); the name of the species; and the count (i.e. how many individuals of the species were observed on the search).

The computation model has 3 distinct steps as illustrated in Figure~\ref{fig:computational-model}.
\begin{enumerate}[noitemsep, topsep=-5pt]
\item We assign observations to a set of stixels (a stixel is represented as a 3D cube).
\item We train a prediction model. Each stixel gets its own prediction model, independent of all computations happening in other stixels..  
\item We average predictions across all stixels that contain a given timestamp and location.
\end{enumerate}

\begin{figure*}
    \includegraphics[width=\textwidth]{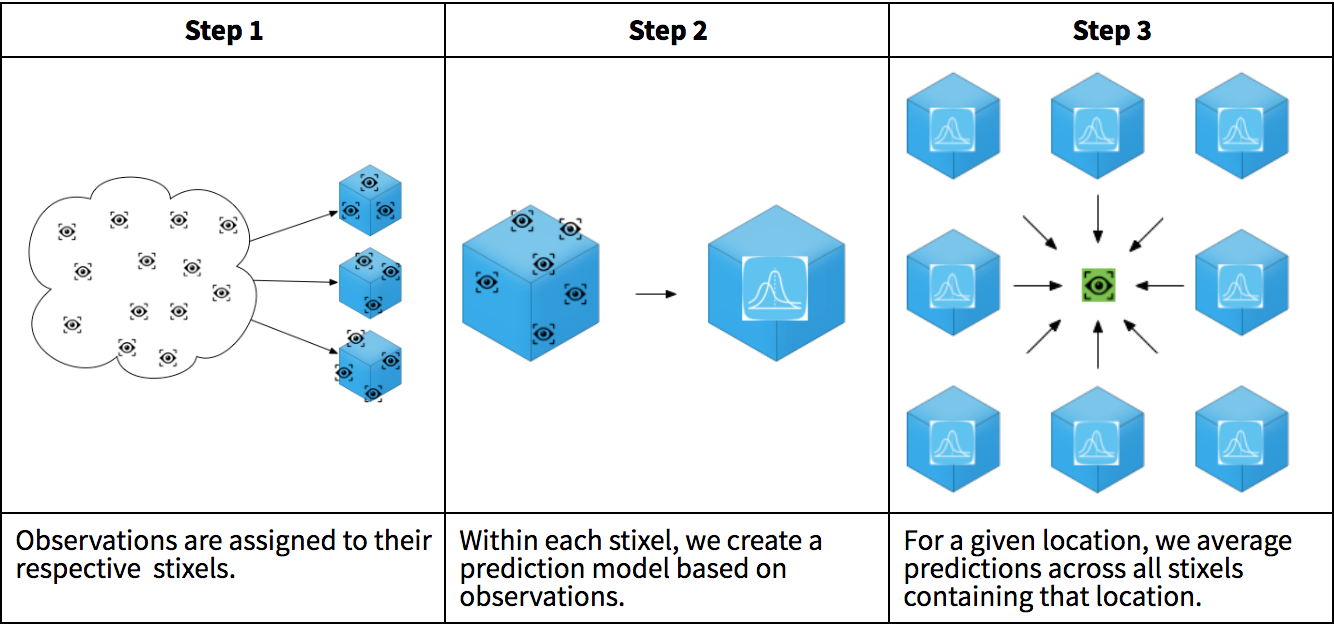}
    \caption{the STEM computation model.}
    \label{fig::computational-model}
  \end{figure*}

This computation model is highly parallel as most of computations occur independently from one another, as illustrated in Figure~\ref{fig::computational-model}.

To deploy and run this computation, the CLO chose to use Spark, an open-source cluster-computing framework for distributing parallel tasks. The team originally used its own high-performance cluster (HPC) to process the data. The data pipeline consists of a Spark launcher written in Python, which calls into 3 scripts written in the R programming language. 

For the STEM analysis of a typical species,  computation involves training models for over 10,000 stixels and making predictions at over 50 million unique space-time locations, consuming an average of 1,600 core hours.

\section{Migrating to the Cloud}\label{sec::migration}
In this section, we describe the various computation deployment we tried. Figure~\ref{fig::cost-comp-1} provides a comparison of the cost for each of them.

\subsection{Running on the high-performance cluster}
CLO originally ran its workflow on a local HPC system. Based on~\cite{Spagnuolo_undated-ih}, we estimate the cost of Equipment + Electricity + Labor + Facilities to \$0.12 core-hour, which translates into \$192 (1,600 core hours $\times$ \$0.12) for a given species.

\subsection{Porting to the cloud}
There are several reasons to migrate to the cloud, as listed in~\cite{noauthor_undated-ne}: flexibility, disaster recovery, automatic software updates, capital expenditure-free, increased collaboration, work from anywhere, document control, security, competitiveness, environmentally friendly.

For CLO, the most relevant reasons were:
\begin{itemize}[noitemsep, topsep=-5pt]
\item Flexibility to scale up and down computations by renting cores 
\item Access to a very competitive infrastructure and related software without having to worry about aging machines
\item No need for extra capital investment: you pay what you use.
\end{itemize}

The team first moved to the Azure Cloud. Using Azure with Spark support (HDInsight) incurs a cost of \$2.46 per 16 core hours (using D14 instances), costing \$246 at current publically advertised rates. The use of the R Server required an extra fee of \$0.08 per core-hour bringing the final cost to \$374.

The team then explored the Amazon cloud, with AWS dedicated instances (aka on-demand) and Spark support via EMR (Elastic Map Reduce) that manages Hadoop and Spark clusters. Using on-demand instances and EMR, the cost was  \$1.33 per 16 core hours, for a total cost of  \$180 for the same computation.

\subsection{Using spot instances}
Cloud offerings presented above assume powerful,  high-availability and dedicated machines for the computation. However, because most STEM computations are not time sensitive, some of these features could be traded for a lower cost.

Amazon Elastic Cloud Computing (EC2) Spot instances are a mechanism that lets you bid on spare Amazon EC2 computing capacity. You make a bid and get available machines. You pay the market price, not your bid. This is similar to second-price auctions. You can use the machine for as long as your bid is above the market price. If your bid goes below the market price, the machine is taken from you, which terminates your computations. 

\begin{figure}[h]
    \centering
    \includegraphics[width=8cm]{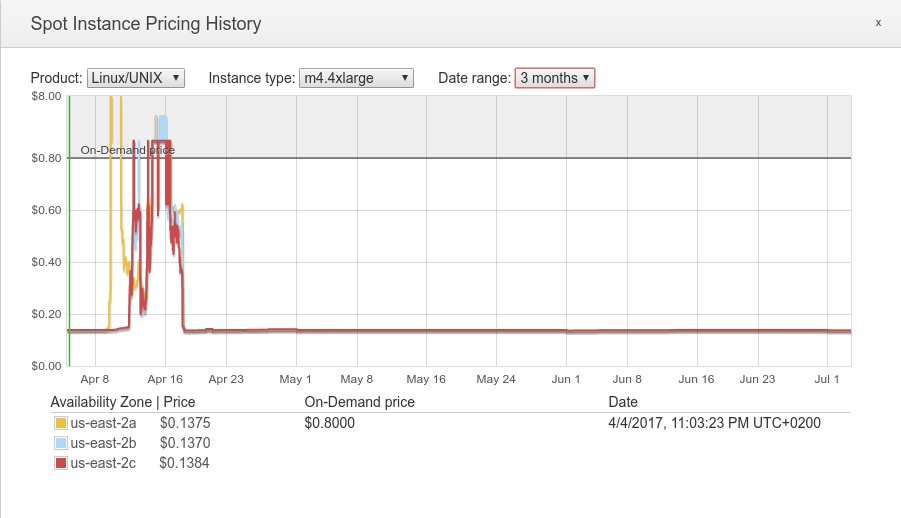}
    \caption{Spot Instance price history.}
    \label{fig::spot-price-history}
\end{figure}

\begin{figure}[h]
    \centering
    \includegraphics[width=6cm]{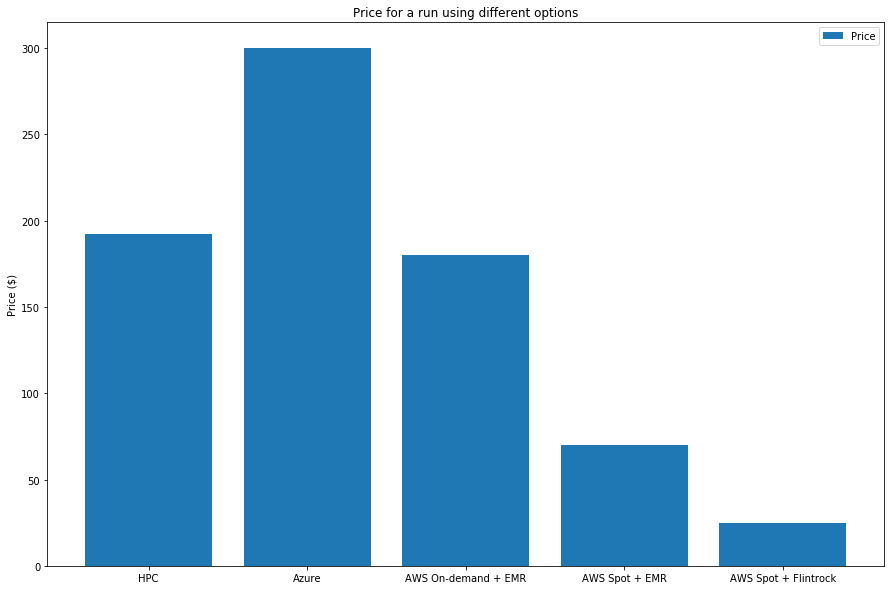}
    \caption{Cost comparison of running the computation.}
    \label{fig::cost-comp-1}
\end{figure}

Spot instances are usually substantially cheaper -- up to an 80\% discount -- compared to on-demand instance prices. Market prices are relatively stable (see Figure~\ref{fig::spot-price-history}), which means that there is no real need for sophisticated bid engineering. EMR also plays well with spot instances.

See Amazon \href{https://aws.amazon.com/ec2/spot/bid-advisor/}{Spot Bid Advisor} for more numbers about savings using spot instances and risks of interruptions.
Using spot instances instead of on-demand instances for most of our computation (we still need some dedicated instances to  monitor the computation itself), we managed to reduce the cost from \$180 to \$70.

\subsection{Using spot instances with Flintrock}
By looking in more details at the cost, we noticed that the cost of EMR was actually higher than the cost of running the spot instances.

\begin{table}[h!]
    \centering
    \begin{tabular}{|l|l|}
    On-demand instances & Spot instances\\ \hline
    \$0.80 - EC2 instance          & \$0.13 - EC2 Spot instance\\
    \$0.24 - EMR                   & \$0.24 - EMR\\
    {\bf EMR $\sim$ 23\% of cost.} & {\bf EMR $\sim$ for 65\% of cost.}
    \end{tabular}
    \caption{Cost of EMR for m4.4xlarge}
    \label{table:cost-EMR}
    \end{table}

EMR offers a very nice service, but the cost is high. What if we could replace EMR?

We would need a solution that can set up a Spark cluster, run a bunch of different bootstrap scripts, and is free. There are two open source solutions corresponding to our needs: \href{https://github.com/amplab/spark-ec2}{Spark EC2} and \href{https://github.com/nchammas/flintrock}{Flintrock}, both command-line tools for launching Apache Spark clusters.

We decided to use Flintrock, because it is a faster and more lightweight version of Spark EC2, with more active development. 
By using Flintrock instead of EMR, the cost is reduced to only \$25 per species.

We summarize the cost of the various options we tried during our journey in the clouds in Figure 4.

\begin{figure}[h]
    \centering
    \includegraphics[width=6cm]{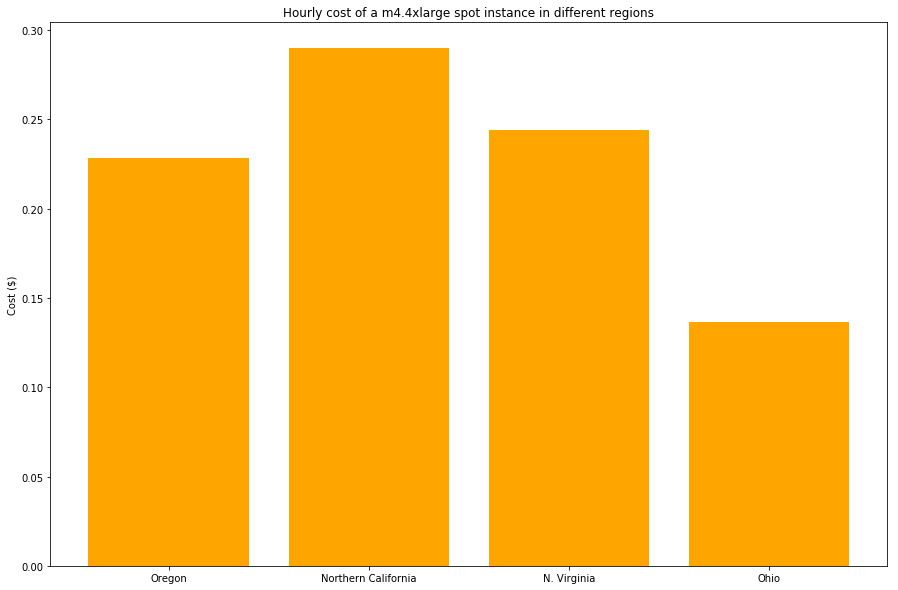}
    \caption{Cost comparison cloud region.}
    \label{fig::cost-comp-2}
\end{figure}

\section{Lessons Learned}\label{sec::lessons}
We now distill the lessons we learned and think are worth sharing, based on our journey.

\subsection*{Lesson 1: the cloud is your friend}
Running computations in the cloud is getting easier and cheaper. Cloud providers provide a large suite of tools to deploy and manage arbitrary computations. See [4] for more details. In our case, the ability to scale up and down, and the availability of spot instances really made a difference.
 
\subsection*{Lesson 2: compare, experiment and shop}
As we saw, not all clouds are created equal. Based on your needs and your level of sophistication, picking the right cloud can save you time and money. This might require spending some time trying a few experiments, but also making sure to understand  the pricing model of the cloud provider you choose to use: computation cost, storage costs, bandwidth costs, costs of auxiliary services you will need, etc. Within a given cloud, there might be some arbitrage to make based on geography, as illustrated in Figure 5.

At the end of the day, make sure you factor all the costs, including the human cost of maintaining and monitoring your computations.

\subsection*{Lesson 3: understand your computation}
The cloud is most useful when your workload is highly parallel. In our case, models could be created independently at the stixel level. Based on this level of parallelism, the team chose to use Spark. However, the use of R reduced some opportunities to fully leverage multicore machines, because some R libraries are not meant to use such machines.

\subsection*{Lesson 4: remember Amdahl's law}
Amdahl's law is a formula used to find the maximum improvement improvement possible by improving a particular part of a system. Four data pipeline, looking at improvements that affect large parts of the computation was essential. The move to spot instances was critical because it reduced by 80\% the cost of all computations. Amdahl's law is also a good checkpoint before embarking on  major code refactoring or even code porting (e.g. from R to a faster language). 

\subsection*{Lesson 5: what do you optimize for}
In the context of this project, the factors to optimize for were clear from the beginning:
The team did not care about speed or availability.
Cost was the number one factor.
But moving away from R (for a faster high-performance language like Python, Julia or C) was not an option: the data science expertise was developed in R and shared inside the R community.

\subsection*{Lesson 6: early optimization is the root of all evil}
Early on in the project, we looked at some obvious shortcomings of the current pipeline: (a) use of R instead of a faster language; (b) use of strings to pass data around as opposed to more compact structures such as protocol buffers; (c) computations that carry both training data and to-be-predicted data.
However, rewriting the pipeline from scratch probably would not have saved 80\% of the computation cost. Similarly, shaving some memory and storage by using more compact encodings would have had very little effect in terms of computation costs, as prices of spot instances  do not vary much based on the size of the instance.

\section{Conclusion}\label{sec::conclusion}
Data is one of the pillars of scientific research. With citizen science and modern technologies such as mobile phones, gathering data has never been easier. Leveraging bird enthusiasts, the eBird project maintains a large dataset of bird observations. But one often forgets that behind a nice bird migration map stands a lot of human work and a complicated data pipeline that is expensive to build and to operate. Very few papers address the cost of producing data and running experiments.

In this paper, we are giving a behind-the-scene view of the eBird data pipeline and our experience migrating the computation from a local cluster to the Cloud. With the Cloud, storage and computation are getting cheaper every day, but they still have cost. Unless you are a large company with infinite resources, operating on a budget means making some trade-offs: reducing the precision of your model, running your model on a subset of the available species.
We describe how, using trial and error and leveraging open source software, we managed to reduce the cost of running the exact same computation by a factor of 6. We also share tips to keep in mind when migrating to the Cloud.

We hope this paper will encourage researchers from the community to share best practices and Cloud providers to offer tools and pricing adapted to this type of research.

\bibliographystyle{ieeetr}
\vspace{2.5mm}
\bibliography{references}

\end{document}